\documentclass[a4paper,12pt]{article}
\usepackage{amsthm}
\usepackage{amsmath}
\usepackage{amssymb}
\usepackage{color}
\usepackage{graphicx}
\usepackage{subfigure}
\usepackage{cite}
\usepackage[linktocpage]{hyperref}
\usepackage{setspace}

\newcommand{\mrd}{\mathrm{d}}

\begin{document}
\let\endtitlepage\relax

\begin{titlepage}
\begin{center}

\newcommand\blfootnote[1]{%
	\begingroup
	\renewcommand\thefootnote{}\footnote{#1}%
	\addtocounter{footnote}{-1}%
	\endgroup
}

\vspace*{-1.0cm}
\renewcommand{\baselinestretch}{1.0}  
\setstretch{1.6}
{\Large{\textbf{Single-rotating Five-dimensional Near-horizon Extremal Geometry in General Relativity}}}

\renewcommand{\baselinestretch}{1.0}  
\setstretch{1.2}

\vspace{8mm}
\centerline{\large{Kamal Hajian\blfootnote{e-mail: kamal.hajian@uni-oldenburg.de}\blfootnote{e-mail: khajian@metu.edu.tr} }}
\vspace{-1mm}
\normalsize
\textit{Institute of Physics, University of Oldenburg, P.O.Box 2503, D-26111 Oldenburg, Germany}\\
\textit{Department of Physics, Middle East Technical University, 06800, Ankara, Turkey}

\renewcommand{\baselinestretch}{1.0}  
\setstretch{1}

\begin{abstract}
	The geometries with SL$(2,\mathbb{R})$ and some axial U$(1)$ isometries are called ``near-horizon extremal geometries" and are found usually, but not necessarily, in the near-horizon limit of the extremal black holes. We present a new member of this family of solutions in five-dimensional Einstein-Hilbert gravity that has only one nonzero angular momentum. In contrast with the single-rotating Myers-Perry extremal black hole and its near-horizon geometry in five dimensions, this solution may have a nonvanishing and finite entropy. Although there is a uniqueness theorem that prohibits the existence of such single-rotating near-horizon geometries in five-dimensional general relativity, this solution has a curvature singularity at one of the poles, which breaks the smoothness conditions in the theorem. 
\end{abstract}
\end{center}
\vspace*{0cm}
\end{titlepage}
\vspace*{-0mm}

\section{Introduction}
Recent observations of black holes \cite{Abbott:2016blz,LIGOScientific:2021djp,Akiyama:2019cqa,EventHorizonTelescope:2022wkp} via electromagnetic and gravitational waves have boosted theoretical research on these mysterious celestial objects. One of their challenging properties is the indication of thermodynamic behavior, such as obeying the four laws of thermodynamics \cite{Bekenstein:1973ft,Hawking:1974rv,Hawking:1976rt,Bardeen:1973gd}. Specifically, a coherent and well-accepted statistical mechanics is missing in the black hole physics community. In spite of many innovative proposals and calculations, the universal identification of black hole microstates is a long-standing question that is still waiting to be resolved. An interesting feature of such an identification would be nonvanishing entropy at zero temperature. This is a general property of extremal black holes, i.e., the holes at zero temperature. 

In practice, if a black hole is more isometric, i.e., if it has more Killing vectors, it is easier to study. In this regard, extremal black holes have a feature that makes them suitable for microstate investigations: their near-horizon geometries enjoy extra isometries than the black holes themselves; the time translation isometry of a stationary extremal black hole is enhanced to an SL$(2,\mathbb{R})$ one. So, the isometries of such a black hole near-horizon geometry with $n$ number of U$(1)$ axial Killings are enhanced to the SL$(2,\mathbb{R})\times$U$(1)^{n}$. 

The near-horizon of the extremal Kerr black hole is one of the simplest examples of the mentioned geometries. It has one U$(1)$ Killing vector, i.e., the axial rotating symmetry generator, to which the angular momentum is associated as a conserved charge. This geometry was first found by Bardeen and Horowitz in 1999 \cite{Bardeen-Horowitz}. Since then, there have been many interesting studies on such near-horizon extremal geometries (NHEG), including microstate counting of extremal black holes (e.g., via Kerr/CFT correspondence \cite{Ent-Funcn-Sen,Kerr/CFT} or symplectic symmetries \cite{Compere:2015bca,Compere:2015mza,Hajian:2017mrf}). 
The classification of these geometries in some gravitational theories and dimensions has been worked out by Kunduri and Lucietti \cite{Kunduri:2008rs,Kunduri:2013gce}. Pedagogical reviews on the classification and explicit examples can be found in \cite{Kunduri:2013gce} and \cite{Compere,Hajian:2015eha} respectively. 

The SL$(2,\mathbb{R})\times$U$(1)^{n}$ isometry, which was alluded to above, is such a restrictive property that it yields uniqueness theorems for the NHEGs in certain dimensions and theories \cite{Kunduri:2007vf,Kunduri:2008rs,Kunduri:2013gce}. However, such uniqueness theorems generically assume some smoothness conditions. As a result, especially when the horizon of the black hole is singular, there can be the possibility of some singular solutions. One of the simplest examples to study such singular horizons is the single-rotating extremal Myers-Perry (MP) black hole in five dimensions \cite{Myers:1986un}. This black hole has an infinitely long bar-shaped horizon with zero area, sometimes called the extremal vanishing horizon (EVH). The near-horizon of EVH, which is obtained by first the single rotation limit and then the near-horizon limit \cite{Fareghbal:2008ar,Fareghbal:2008eh}, has been studied extensively \cite{Sheikh-Jabbari:2011sar,Golchin:2013con,Sadeghian:2014tsa,Sadeghian:2015laa,Sadeghian:2015nfi,Noorbakhsh:2017nde,Sadeghian:2017bpr,Demirchyan:2018lmf}, and it is a static solution without any rotation. The geometry is singular at one of the poles, which will be discussed at the end.   

In this paper, we present a single-rotating NHEG, which is derived by first taking the near-horizon limit of a double-rotating extremal MP black hole and then taking the (appropriately prepared) single-rotation limit. In contrast to the EVH solution, the period of the axial coordinates can be fixed such that the resulting solution has one nonzero angular momentum while obtaining a nonvanishing entropy.  Besides, similar to the near-horizon of the EVH, it has a curvature singularity at one of the poles. This new vacuum solution is the first known single-rotating NHEG in five-dimensional general relativity (putting the trivial NHEG of the extremal Kerr-String solution aside, which is simply the embedding of the extremal Kerr into five dimensions, and is the only single-rotating solution that is allowed by the uniqueness theorem 4.5 in \cite{Kunduri:2013gce}). 

The paper is arranged as follows: In Sec. \ref{NHEG} a short review of NHEGs is provided. In Sec. \ref{solution} the promised single-rotating NHEG solution is presented, and in Sec. \ref{thermo} its thermodynamic properties are investigated. The derivation from the extremal MP black hole and the differences with the near-horizon of the EVH solution are worked out in Secs. \ref{MP} and \ref{EVH} respectively.

\section{Review: near-horizon extremal geometries}\label{NHEG}
Given a Lagrangian density $\mathcal{L}$ in $D$ dimensions, near-horizon extremal geometries are solutions to the equation of motion that have an SL$(2,\mathbb{R})$ and at most a $D-3$ number of U$(1)$ isometries. As their name suggests, these geometries are usually found in the near-horizon limit of extremal black holes. Any NHEG with maximum $D-3$ axial isometry can be written in a coordinate system $(t,r,\theta,\varphi^1,\dots,\varphi^{D-3})$ that makes the SL$(2,\mathbb{R})$ symmetry manifest,
\begin{align}\label{NHEG-metric}
{\mrd s}^2&=\Gamma(\theta)\left[-r^2 \mrd t^2+\frac{\mrd r^2}{r^2}+\alpha\mrd\theta^2+\sum_{i,j=1}^{D-3}\gamma_{ij}(\theta)(\mrd\varphi^i+k^ir\mrd t)(\mrd\varphi^j+k^jr \mrd t)\right],
\end{align}
where
\begin{align}\label{ranges}
t\in (-\infty,+\infty),\qquad r\in \{r<0 \} \text{ or } \{r>0\},\qquad \theta\in[0,\theta_{Max} ], \qquad \varphi^i\sim \varphi^i+2\pi,
\end{align}
$k^i$ are constants over spacetime, and are determined by the equations of motion. The constant $\alpha$ is conventional and determines the domain of the coordinate $\theta$ by fixing the $\theta_{Max}$. The $\Gamma$ and $\gamma_{ij}$ as functions of the coordinate $\theta$ are determined by the equations of motion.

The first two terms in the metric \eqref{NHEG-metric} form an AdS$_2$ in the Poincar\'{e} patch with $r=0$ as the Poincar\'{e} horizon. In these coordinates, the SL$(2,\mathbb{R})\times$U$(1)^{D-3}$ isometry generators are
\begin{equation}
\xi_- =\partial_t\,,\qquad \xi_0=t\partial_t-r\partial_r,\qquad	\xi_+ =\dfrac{1}{2}(t^2+\frac{1}{r^2})\partial_t-tr\partial_r-\frac{1}{r}{k}^i{\partial}_{\varphi^i}, \quad \mathrm{m}_i=\partial_{\varphi^i}
\end{equation}
with the commutation relations 
\begin{align}\label{commutation relation}
[\xi_0,\xi_-]=-\xi_-,\qquad [\xi_0,\xi_+]=\xi_+, \qquad [\xi_-,\xi_+]=\xi_0\,,\qquad [\xi_\mathrm{a},\mathrm{m}_i]=0, \quad \mathrm{a}\in \{-,0,+\}.
\end{align}

\section{Five-dimensional single-rotating solution}\label{solution}
In this section, we present a new five-dimensional NHEG solution to the vacuum gravity with a single angular momentum, which is the main result of this paper. Its derivation from the MP solution will be carried out in Sec. \ref{MP}. Let us consider the Einstein-Hilbert action and Einstein equation in five dimensions, 
\begin{equation}
\mathcal{L}=\frac{1}{16\pi G} R, \qquad R_{\mu\nu}-\frac{1}{2}Rg_{\mu\nu}=0,
\end{equation}
in which $R_{\mu\nu}$ and $R$ are Ricci tensor and scalar respectively. Then, the following single-rotating NHEG in the coordinates $(t,r,\theta,\varphi^1,\varphi^2)$ is a solution.
\begin{align}\label{Solution}
&{\mrd s}^2=\Gamma(\theta)\left[-r^2 \mrd t^2+\frac{\mrd r^2}{r^2}+4\mrd\theta^2+\sum_{i,j=1}^2\gamma_{ij}(\theta)(\mrd\varphi^i+k^ir\mrd t)(\mrd\varphi^j+k^jr\mrd t)\right],\\
& \Gamma= \frac{a^2}{4}\cos^2 \theta, \quad \gamma_{11}=\frac{4\sin^2\theta}{\cos^4\theta}, \quad \gamma_{22}=4, \quad \gamma_{12}=0, \quad k^1=0, \quad k^2=\frac{1}{2}. \label{Solution details}
\end{align}
Explicitly in the more convenient notation $(\varphi^1,\varphi^2)\to (\varphi,\psi)$, it is
\begin{equation}\label{Solution convenient}
\mrd s^2= a^2\cos^2\theta\left(\frac{\mrd r^2 }{4r^2}+  \mrd \theta^2+\frac{\sin^2\theta}{\cos^4\theta}\mrd \varphi^2+\mrd \psi^2+ r\mrd t \mrd \psi\right).
\end{equation}
The coordinate $\theta$ takes the values in $[0,\frac{\pi}{2}]$, while $\varphi\sim \varphi+2\pi$ and $\psi\sim\psi+2\pi$. We have chosen the solution to rotate in the $\psi$ azimuth direction. The case of a single nonzero rotation in the $\varphi$ direction is achieved by the change of the roles of $\varphi \leftrightarrow \psi$ and $\cos\theta \leftrightarrow \sin\theta$ in \eqref{Solution convenient}. The free parameter $a$ is related to the angular momentum of the geometry. One can use different methods for conserved charge calculation in gravity (we used the covariant formulation of charges \cite{Lee:1990gr,Ashtekar:1987hia,Ashtekar:1990gc,Crnkovic:1987at,Wald:1993nt,Iyer:1994ys,Wald:1999wa,Barnich:2001jy,Barnich:2007bf } on the solution phase space \cite{HS:2015xlp}) to find the angular momenta as the charges of the Killing vectors $-\partial_{\varphi}$ and $-\partial_{\psi}$ respectively,
\begin{equation}
J_\varphi=0, \qquad J_\psi=\frac{\pi a^3}{2 G}.
\end{equation}
We observe that the three-dimensional manifold, which is parametrized by the $(t,r,\psi)$ coordinates, is a self-dual orbifold of the AdS$_3$ geometry, which appears in the near horizon of an extremal BTZ black hole \cite{Banados:1992wn}. Such a geometry can also be found in the near horizon of EVHs \cite{Sadeghian:2014tsa,Sadeghian:2015laa,Sadeghian:2015nfi,deBoer:2011zt}. 

Another feature of the metric \eqref{Solution convenient} is that it is singular at $\theta=\frac{\pi}{2}$. To observe this, we can calculate the Kretschmann scalar
\begin{equation}\label{singularity}
R_{\mu\nu\alpha\beta}R^{\mu\nu\alpha\beta}=\frac{72}{a^4\cos^8\theta},
\end{equation}
which diverges if $\theta\to\frac{\pi}{2}$. For the equivalent case of $J_\varphi\neq 0$ and $J_\psi=0$ the singularity is at $\theta\to 0$.  We note that, in spite of the existence of curvature singularity, the solution has finite conserved charges and satisfies the NHEG thermodynamic relations, which will be discussed in the next sections.

\section{Thermodynamic properties}\label{thermo}
For a generic thermodynamic system, the first law of thermodynamics is a universal law, i.e., independent of the details of the system,  which makes a relation between the variation of the thermodynamic properties of the system at nonzero temperatures\cite{Bardeen:1973gd}. At zero temperature, most thermodynamic systems have zero entropy, and the only possibility for the variation of the entropy is to excite the system to have a nonzero temperature, and this is governed by the expansion of the first law around zero temperature. However, in black hole thermodynamics, there is the possibility of nonvanishing entropy at zero temperature. So, it is possible to  change the entropy without changing the vanishing temperature, i.e., by changing one extremal black hole to the adjacent extremal one. In this case, the first law (neither its standard form at nonzero temperature, nor its expansion around $T=0$ without implementing an extra assumption \cite{Johnstone:2013ioa}) does not capture the behavior of $\delta S$, simply because by $T=0$ and $\delta S \neq 0$ the $T\delta S$ term disappears from this law. As a result, the remaining parts of the first law would be just the extremality condition between the mass and other charges of the black hole without relation to the entropy.      

In parallel with the standard first law and Smarr formula, the extremal black holes enjoy two extra relations that govern the $\delta S$ and $S$. These relations have been found and proved in the context of NHEGs \cite{Hajian:2013lna,Hajian:2014twa} (following the pioneering work \cite{Ent-Funcn-Sen}). The essential point to derive these extra relations is to note that although an NHEG does not have an event horizon, it admits an infinite number of Killing horizons with identical thermodynamic properties. Explicitly, any surface of constant $t$ and $r$, calling it $\mathcal{H}$ with $t=t_{_\mathcal{H}}$ and $r=r_{_\mathcal{H}}$, is the bifurcation of the two null surfaces
\begin{equation}\label{NHEG horizon}
t+\frac{1}{r}=t_{_\mathcal{H}}+\frac{1}{r_{_\mathcal{H}}}, \qquad t-\frac{1}{r}=t_{_\mathcal{H}}-\frac{1}{r_{_\mathcal{H}}},
\end{equation}
with the following Killing vector as the generator of the horizon  
\begin{equation}
\zeta_{_\mathcal{H}}=n^\mathrm{a}_{{_\mathcal{H}}} \, \xi_\mathrm{a}+k^i \mathrm{m}_i.
\end{equation}
In the definition of $\zeta_{_\mathcal{H}}$ summation over $\mathrm{a}\in \{-,0,+\}$ and $i=\{1,2,..,D-3\}$ is understood, and the $n^\mathrm{a}$ is 
\begin{equation}
n^-=-\frac{t^2r^2-1}{2r}, \qquad n^0=tr, \qquad n^+=-r.
\end{equation} 
Geometrically, $n^\mathrm{a}$ is the unit vector from the center in a $3$-dimensional flat space to the $(t,r)$ point of the AdS$_2$ geometry that is immersed in it (see the details in Sec. 5 of \cite{Hajian:2015eha}).  The Killing vector $\zeta_{_\mathcal{H}}$ is null on the horizon in \eqref{NHEG horizon} and vanishes at the bifurcation $\mathcal{H}$. Similar to the celebrated Wald entropy formulation\cite{Wald:1993nt,Iyer:1994ys}, the NHEG entropy is defined as the conserved charge of  this Killing vector calculated on the bifurcation surface $\mathcal{H}$ \cite{Hajian:2013lna}. For Einstein-Hilbert gravity, the result turns out to be the Bekenstein-Hawking entropy\cite{Bekenstein:1973ft} 
\begin{equation}
S=\frac{A_{_\mathcal{H}}}{4G}.
\end{equation}
Thanks to the SL$(2,\mathbb{R})$ isometry, $\mathcal{H}$ can be  any one of the surfaces of the $(t,r)$ constant with the same result for the $S$. With a well-defined entropy for the NHEGs and following similar steps as in Wald's proof of the first law in \cite{Iyer:1994ys}, the NHEG thermodynamic laws are found, which for the vacuum solutions they are
\begin{align}\label{NHEG laws}
\frac{\delta S}{2\pi}=  k^i \delta J_i, \quad \qquad \text{and} \quad \qquad \frac{S}{2\pi}=  k^i J_i.
\end{align}

The NHEG solution in \eqref{Solution} and \eqref{Solution details} satisfies these two relations; one can use different methods for conserved charge calculation in gravity (we used the covariant formulation of charges \cite{Ashtekar:1987hia,Ashtekar:1990gc,Crnkovic:1987at,Wald:1993nt,Iyer:1994ys,Wald:1999wa,Barnich:2001jy,Barnich:2007bf } on solution phase space \cite{HS:2015xlp}) to find its entropy and angular momenta as charges of $\zeta_{_\mathcal{H}}$, $-\mathrm{m}_1=-\partial_{\varphi^1}$ and $-\mathrm{m}_2=-\partial_{\varphi^2}$,
\begin{equation}\label{Solution charges}
S=\frac{\pi^2 a^3}{2G}, \qquad J_1=0, \qquad J_2=\frac{\pi a^3}{2 G}.
\end{equation}
By the $k^1=0$ and $k^2=\frac{1}{2}$ the NHEG thermodynamic laws  \eqref{NHEG laws} are satisfied.

\section{Derivation from extremal Myers-Perry black hole}\label{MP}
Analogous black holes to the Kerr in higher dimensions are called Myers-Perry black holes. The five-dimensional MP solution is characterized by its mass and two angular momenta, with the metric 
\begin{align}\label{MP metric}
\mathrm{d}s^2= -&(\frac{-\Delta+a^2\sin^2\theta+b^2\cos^2\theta+\frac{a^2b^2}{{\hat{r}}^2}}{\rho^2})\mathrm{d}{\hat{t}}^2+\frac{\rho ^2}{\Delta}\mathrm{d}\hat {\hat{r}}^2+\rho^2\mathrm{d}\theta ^2 \nonumber\\
&+2\left(\Delta-({\hat{r}}^2+a^2)-\frac{b^2({\hat{r}}^2+a^2)}{{\hat{r}}^2}\right)\frac{a\sin^2\theta}{\rho^2}\,\mathrm{d}t\, \mathrm{d}\chi^1\nonumber\\
&+2\left(\Delta-({\hat{r}}^2+b^2)-\frac{a^2({\hat{r}}^2+b^2)}{{\hat{r}}^2}\right)\frac{b\cos^2\theta}{\rho^2}\,\mathrm{d}t\, \mathrm{d}\chi^2\nonumber\\
&+\left(-\Delta a^2\sin^2\theta+({\hat{r}}^2+a^2)^2+\frac{b^2({\hat{r}}^2+a^2)^2\sin^2\theta}{{\hat{r}}^2}\right)\frac{\sin^2\theta}{\rho^2}\,\mathrm{d}\chi^1 \mathrm{d}\chi^1\nonumber\\
&+\left(-\Delta b^2\cos^2\theta+({\hat{r}}^2+b^2)^2+\frac{a^2({\hat{r}}^2+b^2)^2\cos^2\theta}{{\hat{r}}^2}\right)\frac{\cos^2\theta}{\rho^2}\,\mathrm{d}\chi^2 \mathrm{d}\chi^2\nonumber\\
&+2\left(-\Delta +\frac{({\hat{r}}^2+a^2)({\hat{r}}^2+b^2)}{{\hat{r}}^2}\right)\frac{ab\sin^2\theta\cos^2\theta}{\rho^2}\,\mathrm{d}\chi^1 \mathrm{d}\chi^2\,,
\end{align}
where
\begin{align}
\rho^2 \equiv {\hat{r}}^2+a^2 \cos^2 \theta+b^2\sin^2\theta, \qquad 
\Delta \equiv \frac{({\hat{r}}^2+a^2)({\hat{r}}^2+b^2)}{{\hat{r}}^2}+2m\,.
\end{align}
The range of the spherical coordinates is $\theta\in [0,\frac{\pi}{2}]$ and $\chi^i\in[0,2\pi]$. There are three free parameters $m$, $a$ and $b$ in this metric. The extremality constrains the mass to be a function of the angular momenta in the solution via the constraint $2m=(a+b)^2$. The near-horizon of the extremal MP black hole is achieved by the coordinate transformation 
\begin{align}
\hat t=\frac{\alpha_0 r_h t}{\epsilon}, \qquad \hat r=r_h(1+\epsilon r), \qquad \chi^i=\varphi^i+\frac{\Omega^i \alpha_0 r_h t}{\epsilon}, 
\end{align}  
in which
\begin{equation}
r_h=\sqrt{ab},\qquad  \alpha_0=\frac{(a+b)^2}{4r_h^2}, \qquad \Omega^1=\Omega^2=\frac{1}{(a+b)},
\end{equation}
followed by the limit $\epsilon\to 0$. The NHEG that is found has the general form of \eqref{Solution} with
\begin{align}
&\Gamma=\frac{1}{4}(a+b)(a\cos^2\theta+b\sin^2\theta), \qquad  k^1=\frac{1}{2}\sqrt{\frac{b}{a}}, \qquad k^2=\frac{1}{2}\sqrt{\frac{a}{b}},\nonumber \\
&\gamma_{ij}=\dfrac{4}{(a\cos^2\theta+b\sin^2\theta)^2}
\begin{pmatrix}
 a (a+b\sin^2\theta) \sin ^2\theta &  a b \cos ^2\theta \sin ^2\theta \\ \ \ & \ \ \\
 a b \cos ^2\theta \sin ^2\theta &  b (b+a\cos^2\theta)\cos ^2 \theta\\
\end{pmatrix}.\label{MP NHEG}
\end{align}
The entropy and angular momenta of this solution are functions of the two free parameters $a$ and $b$, 
\begin{equation}\label{MP SJ}
 \frac{S}{2\pi}=\frac{\pi \sqrt{ab} (a+b)^2}{4G}, \qquad J_1=\frac{\pi a (a+b)^2}{4G}, \qquad J_2=\frac{\pi b (a+b)^2}{4G},
\end{equation}
which coincide with the entropy and angular momenta of the initial extremal MP black hole and satisfy the NHEG laws in \eqref{NHEG laws}. However, to request one of the angular momenta to vanish, $a$ or $b$ should be set to zero. In this limit, not only one of the angular momenta vanishes, but also the entropy is zero, as expected from the original extremal MP. Nonetheless, such a limit is not well behaved because it makes some of the metric components in \eqref{NHEG-metric} with \eqref{MP NHEG} to blow up.

Although the limiting procedure above is not successful in creating a single-rotating NHEG, it can be modified such that it yields a new solution with different conserved charges, which is the solution in Eqs. \eqref{Solution} and \eqref{Solution details}. Let us assume that we want $b\to 0$. To this end, by redefinition of the coordinate 
\begin{equation}\label{redefinition}
\varphi^2\to \frac{\varphi^2}{2k^2},
\end{equation}
and then taking the limit $b\to 0$ in \eqref{MP NHEG}, the single-rotating solution is Sec. \ref{solution} is found. The period of the new axial coordinate [which corresponds to the $\psi$ in \eqref{Solution convenient}], can be set to be $2\pi$ in order to keep the nonvanishing angular momentum intact, i.e., to be equal to $\frac{\pi a^3}{2 G}$. However, noting the singularity at the $\theta=\frac{\pi}{2}$, this period is an arbitrary parameter. Especially, if the solution is considered to be strictly derived from the MP black hole, then by the limit $b\to 0$ the constant $k^2$ in the Eq. \eqref{redefinition} diverges and admits a vanishing period for this axial direction.  

Notice that the solution \eqref{Solution} and \eqref{Solution details}, which is found by the limits above and adjustment of the periods, has different properties in comparison with the NHEG of MP. Specifically in $b\to 0$ limit, the $J_2$ remains finite and nonzero, while the $J_1$ vanishes as in \eqref{Solution charges}, which is the reverse of the extremal MP angular momenta in \eqref{MP SJ}. Moreover, the entropy is finite and nonzero, and so it is different from the zero entropy of the single-rotating extremal MP case in \eqref{MP SJ}.

\section{Comparison with near-horizon of EVH solution}\label{EVH}
If one of the two angular momenta of the extremal MP black hole is set to zero, then the horizon becomes singular in the shape of an infinitely long bar with zero area. Such an extremal geometry with a vanishing area of their horizon is called an extremal vanishing horizon (EVH) whose entropy vanishes. The near-horizon of such EVHs has been studied first in \cite{Bardeen-Horowitz} and investigated more in detail in Refs. \cite{Sheikh-Jabbari:2011sar,Golchin:2013con,Sadeghian:2014tsa,Sadeghian:2015laa,Sadeghian:2015nfi,Noorbakhsh:2017nde,Sadeghian:2017bpr,Demirchyan:2018lmf}. To see how it is different from the single-rotating solution in this paper, we review its derivation. To this end, we can put $a$ or $b$ in Eq. \eqref{MP metric} equal to zero and then take the near-horizon limit. Conventionally, if we set $b=0$, the near-horizon limit of the EVH is taken by the transformation
\begin{align}
\hat t=\frac{t}{\epsilon}, \qquad \hat r=\epsilon r, \qquad \chi^1=\varphi^1+\frac{\Omega^1 t}{\epsilon}, \qquad \chi^2=\frac{\varphi^2}{\epsilon},
\end{align}
followed by the $\epsilon\to 0$ limit \cite{Bardeen-Horowitz}. The result is 
\begin{align}\label{NHEVH metric}
\mrd s^2=\cos^2\theta\, \left(-\frac{r^2}{a^2}\mrd t^2+\frac{a^2}{r^2}\mrd r^2+r^2 (\mrd\varphi^2)^2\right)+a^2\cos^2\theta \mrd \theta^2+a^2\tan^2\theta(\mrd\varphi^1)^2.
\end{align}
This metric has a (pinching $\varphi^2 \sim \varphi^2+2\pi\epsilon $) AdS$_3$ sector in the parenthesis, is static, and has zero angular momenta. The latter makes the metric \eqref{NHEVH metric} different from our new solution in Sec. \ref{solution}. However, it has the same singularity at $\theta\to\frac{\pi}{2}$, i.e., the Kretschmann scalar diverges exactly as the one in \eqref{singularity}. In this respect, the two solutions share similar behaviors. 

It is also worth emphasizing that the EVH limit keeps the entropy of the original extremal black hole intact, i.e., zero, while killing its nonvanishing angular momentum. In contrast, in the solution presented in Sec. \ref{solution} there is also a possibility for the reverse: keeping the angular momentum intact while turning the entropy on. 

\section{Conclusion}
In this work, we presented the first nontrivial single-rotating NHEG in five-dimensional vacuum theory. Although the solution suffers from a curvature singularity, it has well-defined conserved charges, and satisfies NHEG thermodynamic relations. In addition, as a new feature, it has a self-dual orbifold of AdS$_3$ in a part of the metric.  We also showed that one way to derive this solution is to take the (adjusted) single-rotational limit of the NHEG of the MP black hole. However, the solution may have different entropy and angular momenta in comparison with the original single-rotating extremal black hole. Consequently, if independent physical conditions are applied to fix the period, then it can be considered an independent solution. 

Investigation of the global properties of this solution, especially studying the singularities and global Killing vectors, are interesting parts of the future analysis. Besides, the generalization of such an analysis to higher-dimensional EVHs as well as studying the pinching situation of the adjusted axial direction will be postponed for future works. 

\noindent \textbf{Acknowledgments:} I am very thankful for the kind support from Jutta Kunz and Bayram Tekin in Oldenburg University and METU. I would also like to thank them, as well as Eugen Radu, Shahin Sheikh-Jabbari, and Mohammad H. Vahidinia for useful discussions and comments. This work has been supported by TÜBITAK international researchers program No. 2221.

\end{document}